\documentclass[aps,prl,twocolumn,superscriptaddress,nofootinbib,longbibliography,10pt]{revtex4-2}
\usepackage{etoolbox}
\usepackage{dcolumn}
\usepackage{amsmath,amssymb,amsfonts,mathtools}
\usepackage{mathrsfs,bbold}
\usepackage{graphicx}
\usepackage[colorlinks=true,urlcolor=blue,citecolor=red,linkcolor=blue]{hyperref}
\usepackage{accents}
\newlength{\dhatheight}
\usepackage{natbib}
\usepackage{wrapfig}
\usepackage{flushend,BOONDOX-cal,BOONDOX-frak}

\makeatletter
\newsavebox{\@brx}
\newcommand{\llangle}[1][]{\savebox{\@brx}{\(\m@th{#1\langle}\)}%
  \mathopen{\copy\@brx\kern-0.5\wd\@brx\usebox{\@brx}}}
\newcommand{\rrangle}[1][]{\savebox{\@brx}{\(\m@th{#1\rangle}\)}%
  \mathclose{\copy\@brx\kern-0.5\wd\@brx\usebox{\@brx}}}
\makeatother

\begin{document}
\title{`Seeing' the quantum ripples of spacetime}
\author{Soham Sen}
\email{sensohomhary@gmail.com}
\affiliation{Department of Astrophysics and High Energy Physics, S. N. Bose National Centre for Basic Sciences, JD Block, Sector-III, Salt Lake City, Kolkata-700 106, India}
\author{Vlatko Vedral}
\email{vlatko.vedral@physics.ox.ac.uk}
\affiliation{Clarendon Laboratory, University of Oxford, Park Road, Oxford OX1 3PU, United Kingdom}

\begin{abstract}
\noindent We propose a novel way of detecting gravitons using emission of photons from charged array of quantum harmonic oscillators placed inside of a cavity while the cavity is being pumped with low frequency photons. We observe that when the detector is in its ground state, a single graviton is absorbed by the detector while it jumps a single energy level by simultaneously emitting a photon. We also observe that while the detector de-excites from an higher energy level, it spontaneously emits a high frequency graviton, by absorbing a single photon. This analytical outcome encourages us to propose a very simple tabletop  graviton detector model as the transition probabilities can be significantly enhanced by pumping photons in the initial state of the system. This mechanism gives us a physical way to `visualize' the effect of gravitons with a relativistic system. We also show that Dyson's original argument on the non-detectability of gravitons can be completely circumvented using our proposed graviton detector.
\end{abstract}
\maketitle
\noindent \textit{\textbf{Introduction:}} Recent advances in low energy quantum field theory of gravity aims primarily towards investigating phenomenological aspects of the interaction of graviton with quantum matter. A series of works \cite{MarlettoVedral2,
EntanglementQuantumGravity,
EntanglementQuantumGravity2,
EntanglementQuantumGravity3,
QGravLett,QGravD,KannoSodaTokuda,KannoSodaTokuda2}
have investigated the response of quantum matter towards incoming quantum gravitational fluctuations. It has been thoroughly investigated in the works \cite{QGravLett,QGravD,KannoSodaTokuda,KannoSodaTokuda2}, the effects of graviton induced noise fluctuation on quantum matter and in \cite{KannoSodaTokuda,KannoSodaTokuda2}, it was found that the noise induced by graviton lead to decoherence effects in entangled relativistic set-ups leading to a future possibility of the detection of graviton induced noise in advanced interferometric detectors. This analyses primarily look for the effective change in the Langevin-like stochastic differential equation of motion. A more direct approach to detect single gravitons was proposed in \cite{Pikovski} and it was shown that through continuous sensing of quantum jumps one can resolve stimulated absorption of gravitons. The research also emphasises on an experimental proposal that depicts single graviton detection as detectable in advanced experimental set up. In a more analytically focussed approach \cite{OTMGraviton}, a straightforward representation of a resonant gravitational wave detector was taken where it was observed that the stimulated absorption of gravitons by the detector can be mapped exactly to the semiclassical analogue, however, the gravitons truly show a spontaneous emission signature which is significantly weaker than the stimulated emission scenario. 

\noindent A very interesting phenomena, however, is possible to observe in nature where in the presence of a static magnetic field electromagnetic and gravitational waves can interchange into each other which is famously known as the Gertsenshtein effect \cite{Gertsenshtein}. In these experimental set ups a photon-shielded cavity is  filled with a magnetic field where the gravitational waves enter the cavity allowing for graviton-photon conversion. There are several analyses \cite{Polzanin,PairConversion, JiroSoda,HwangNoh,QuGrav,Germain} focussing on the photon-graviton conversion mechanism and emphasising on experimental set-ups for graviton detection using this mechanism \cite{UpperLimit,QuGrav,Schutzhold}. In \cite{Schutzhold}, it was analytically found that the detection of the stimulated emission or absorption of gravitons by an optical Weber bar-like is possible using current experimental methods.

\noindent In this work, we go a step further to the already proposed Weber bar models in \cite{Pikovski,OTMGraviton} by considering a conducting Weber bar gravitational wave detector carrying a finite charge in the presence of a dynamical electromagnetic field. The charge allows coupling of the detector with the external dynamical electromagnetic field and we have allowed minimal coupling for our analysis. The entire set-up is considered to a be placed inside of a cavity which is electromagnetically shielded to prevent interaction with any external electromagnetic fluctuations and the entire set up is then considered to be perturbed by a background gravitational fluctuations where the fluctuations are quantized. This is a direct quantum gravitational extension of our previous work \cite{OTM}. This set-up allows for a graviton-photon-detector coupling where the strength of the coupling constant can be controlled by tuning the finite charge of the Weber bar. Apart from the two-party interaction including the graviton-detector, photon-detector, and graviton-photon couplings our main focus is to investigate this three-party coupling term and the underlying physics. It is quite evident that the graviton-photon coupling term will result in the well investigate quantum Gertsenshtein effect, however, the physics of the three-particle term is not still well investigated in the literature. In this analysis, we primarily investigate the three physical scenarios, (1) the conversion of gravitons into photons and vice-versa when the detector excites from a lower energy level, (2) the conversion of photons into gravitons and vice-versa when the detector de-excites from an higher excited state, and (3) spontaneous emission scenario in case of single photons and gravitons and their detection mechanism if any of the above processes are viable and novel physically as well as experimentally. 


\noindent \textit{\textbf{The model:}} We start with the basic background for the model which is considered to be a flat Minkowski background with small spacetime fluctuations given as $g_{\mu\nu}=\eta_{\mu\nu}+h_{\mu\nu}$.
There will be three parts of the action, the first one being the geometric part, the second part being the electromagnetic part of the action, and the final part being the the matter or detector part of the action.
The geometric part of the action will be given by the Einstein-Hilbert action and the transverse-traceless gauge condition will be implemented for getting rid of any redundant unphysical degrees of freedom. We conceptualize the detector as a two-mass system one needs to consider one of the ends of the detector to follow a time-like geodesic. This end is considered to be an object with a heavier mass, whereas the collective phonon modes is considered as a particle with smaller mass $m_0$ such that $m_0\gg m_\infty$ where the mass $m_0$ is connected to $m_\infty$ by a spring with oscillation frequency $\omega_0$. The coordinate of $m_0$ can simply be represented by the Fermi-normal coordinates $\mathcal{Y}^\mu=\{t,\xi^i\}$, with $\xi^i$ denoting the spatial part of the geodesic separation in the Fermi-normal coordinates. Finally, for the electromagnetic part we consider the Maxwell action in curved spacetime where the Coulomb gauge condition is implemented ($A^0=0$, and $\vec{\nabla}\cdot\vec{A}=0$). The complete action for the model system reads $S
=\frac{1}{16\pi G}\int d^4 x\sqrt{-g}R-\frac{1}{4}\int d^4x \sqrt{-g}F_{\mu\nu}F^{\mu\nu}-m_0\int dt \left[\sqrt{-g_{\mu\nu}\dot{\mathcal{Y}}^\mu\dot{\mathcal{Y}}^\nu}+\frac{\omega_0^2}{2} \mathcal{Y}_\mu\mathcal{Y}^\mu-\frac{q}{m_0} g_{\mu\nu}A^\mu \dot{\mathcal{Y}}^\nu\right]$
with $m_0$ being the mass of the entire detector modelled by a single particle carrying the effective  mass of the phonon mode with a mode frequency of $\omega_0$ and $q$ being the overall charge carried by the detector. It is important to note that the final term in the last line of the action in the above equation actually gives the EM-detector coupling in curved background. 

\noindent \textit{Quantizing the model system:} We shall now proceed towards the quantization of model system. To quantize, at first, we implement a discrete mode decomposition of the gravitational fluctuation as well as the electromagnetic gauge field. In the transverse-traceless gauge the gravitational perturbation can be decomposed as
\begin{equation}\label{II.3}
h^{\text{TT}}_{\mu\nu}(t,\vec{x})\rightarrow\bar{h}_{ij}(t,\vec{x})=\frac{1}{\sqrt{\hbar G}}\sum_{\vec{k},s}h_{s}(t,\vec{k})e^{i\vec{k}\cdot\vec{x}}\epsilon_{ij}^s(\vec{k})
\end{equation}
with $h^s(t,\vec{k})$ denoting the Fourier mode functions, $\epsilon^s_{ij}(\vec{k})$ denoting the polarization tensor, and $s=\{+,\times\}$ for the system being in the transverse-traceless gauge. The above mode decomposition helps us to obtain the analytical form of the Einstein-Hilbert action in the transverse-traceless gauge up to second order in the perturbation term as $S_{\text{EH}}=\frac{m}{2}\int dt\sum_{\vec{k},s}(|\dot{h}_s(t,\vec{k})|^2-k^2|h_s(t,\vec{k})|^2)$ where $m\equiv \frac{V}{16\pi \hbar G^2}$ and we have made use of the identity $\int d^3x~ e^{i(\vec{k}+\vec{k}')\cdot \vec{x}}=V\delta_{\vec{k},-\vec{k}'}$ and .  Here, $V$ denotes the quantization volume of the gravitational wave. For the vector-field, we consider the mode decomposition in the Coulomb gauge to be of the form given as
\begin{equation}\label{II.4}
A_{i}(t,\vec{x})=\frac{1}{\sqrt{\hbar G^2}}\sum_{\vec{k}_P,P}A_{P}(t,\vec{k}_P)e^{i\vec{k}_P\cdot\vec{x}}\epsilon^s_{i}(\vec{k}_P)
\end{equation}
with $A_{P}(t,\vec{k}_P)$ denoting the mode function in the Fourier space and $\epsilon^s_i(\vec{k}_P)$ denoting the electromagnetic polarization tensor. The above mode decomposition helps us to write down the base part of the Maxwell action as
$S_{\text{M}}^{(0)}=\frac{m_P}{2}\int dt\sum_{\vec{k}_P,P}(|\dot{A}_P(t,\vec{k}_P)|^2-k_P^2|A_P(t,\vec{k}_P)|^2)$ where $m_P\equiv \frac{V}{\hbar G^2}=16
\pi m$. Both $m$ and $m_P$ have the dimension of mass which gives the same dimension for the Fourier mode functions $h_s(t,\vec{k})$ and $A_P(t,\vec{k}_P)$. In order to arrive at the final simplified structure of the Hamiltonian we need to provide some physical inputs. At first, we can set $\bar{h}_{ij}(t,\vec{x})=\bar{h}^*_{ij}(t,\vec{x})$ as the gravitational perturbations are real and one can also impose the similar condition for the vector field $A_{i}(t,\vec{x})=A^*_i(t,\vec{x})$. Secondly, a more simpler model, we can consider that the propagation vectors for the gravitational wave as well as the electromagnetic wave to be parallel, that is  $\vec{k}\parallel\vec{k}_P$. Finally, a resonant bar is generally considered as a long metallic rod with comparatively smaller height and width. Hence, we can effectively express the geodesic deviation as $\xi^i=\{\xi,\varepsilon_y,\varepsilon_z\}$ and neglect $\mathcal{O}(\varepsilon^2)$ as well as any dynamical contributions form the directions perpendicular to $\xi$ ($\mathcal{O}(\dot{\varepsilon})$ or higher).
We consider the direction of propagation for the gravitational wave as well as the electromagnetic wave to be the $z$-direction which gives us $\vec{k}=\{0,0,k\}$ and $\vec{k}_P=\{0,0,k_P\}$.
We also demand that the gravitational wave carries plus polarization only and the same goes for the electromagnetic wave. Restricting to single mode consideration only, we can write down $\Re[h_{s}(t,k)]=h(t)$ and $\Im[h_{s}(t,k)]=\bar{h}(t)$ and $\Re[A_{P}(t,k_P)]=A(t)$ (considering the Fourier mode function for the gauge field to be completely real). This helps us to write down the action for the entire model system in a surprisingly simple form and from there one can read of the Lagrangian. We can consider that the single-mode Fourier mode function for the gravitational wave has no complex counter part which sets $\bar{h}$ to zero we get rid of all $p_{\bar{h}}$ contributions as well which allows us to write down the Hamiltonian of the model system in a much simpler form as (where all higher order terms have been dropped)
\begin{equation}\label{II.5}
\begin{split}
H\simeq&H_0+\frac{2g_P h}{m_P}\left(\frac{p_A^2}{2m_P}-\frac{1}{2}m_p\omega_P^2A^2\right)\\
+&\frac{g_hp_h}{2mm_0}(\xi\pi_\xi+\pi_\xi \xi)-\frac{q_P}{m_0}A\pi_\xi-\frac{g_hq_P}{mm_0}p_hA\xi~.
\end{split}
\end{equation} 
with $g_h\equiv \frac{m_0}{2\sqrt{\hbar G}}$, $g_P\equiv\frac{m_P}{2\sqrt{\hbar G}}$, $q_P\equiv \frac{q}{\sqrt{\hbar G^2}}$, and $H_0$ giving the base part of the action,

\noindent We are now in a position to quantize the entire model system. We now raise the phase space variables to operator status and implement suitable canonical commutation relations as
$[\hat{h},\hat{p}_{h}]=[\hat{A},\hat{p}_A]=[\hat{\xi},\hat{\pi}_\xi]=i\hbar$.  Here, $\{\hat{b},\hat{b}^\dagger\}$ denote the graviton annihilation and creation operator and $\{\hat{a},\hat{a}^\dagger\}$ denoting the photon annihilation and creation operators, however, $\{\hat{\chi},\hat{\chi}^\dagger\}$ denote the lowering and raising operators for the energy level of the detectors. One can also consider them as phonon creation and annihilation operators. It is important to observe that the frequency of the photons observe a shift and we denote this shifted frequency by $\bar{\omega}_P$ where $\bar{\omega}_P^2\equiv \omega_P^2+\frac{q_P^2}{m_0m_P}$. 
We need to now methodically look at each terms of the interaction Hamiltonian. The operatorial form of the first term in the right hand side of eq.(\ref{II.5}) after the base Hamiltonian $H_0$ actually gives the photon-detector conversion term which says that if the graviton couples to the photons (depending on the strength of the graviton-photon coupling term $g_P$), then a graviton will exchange into photons and vice-versa however, the detector remains unaffected in this process. The next term, that is $\frac{g_h}{2mm_0}\hat{p}_h\otimes \hat{\mathbb{1}}_{\text{P}}\otimes (\hat{\xi}\hat{\pi}_\xi+\hat{\pi}_\xi\hat{\xi})$ gives the usual graviton-detector coupling term which we have already investigated in \cite{OTMGraviton}. The penultimate term $-\frac{q_P}{m_0}\hat{\mathbb{1}}_{\text{GW}}\otimes \hat{A}\otimes\hat{\pi}_\xi$ gives the photon-detector coupling term which will result in the excitation and de-excitation of the detector resulting in resonant absorption and spontaneous emission of photons.  In our analysis, however, the most interesting physics will come from the final term in the right hand side of eq.(\ref{II.5}) resulting in a graviton-photon-detector coupling. We are most interested in this trilinear interaction Hamiltonian given by $\hat{\mathcal{H}}_{\text{int}}=-\frac{g_hq_P}{mm_0}\hat{p}_h\otimes\hat{A}\otimes\hat{\xi}$ and investigate the physics coming in such an analytical scenario with the possibility of experimental detectability. 

\noindent \textit{\textbf{Graviton to photon conversion by the detector:}} To calculate the unitary time evolution operator, we need to write down the interaction Hamiltonian in the interaction picture. 
In this picture all the phase space operators gets replaced by its interaction picture analogue. The three-mode interaction Hamiltonian in the interaction picture reads 
\begin{equation}\label{III.6}
\begin{split}
\hat{\mathcal{H}}_{int}^I=e^{\frac{i}{\hbar}\hat{H}_0 t}\hat{\mathcal{H}}_{int}e^{-\frac{i}{\hbar}\hat{H}_0 t}=&-\frac{g_hq_P}{mm_0}\hat{p}_h^I\otimes \hat{A}^I\otimes\hat{\xi}^I~.
\end{split}
\end{equation}
The unitary time-evolution operator then can be expressed in terms of the above interaction Hamiltonian in the interaction picture as $\hat{\mathcal{U}}^I(t,t_i)=\mathcal{T}\left[e^{-\frac{i}{\hbar}\int_{t_i}^t dt' \hat{\mathcal{H}}_{int}^I(t')}\right]$.
As we are considering the dimensionless coupling coefficients in $\hat{H}_{int}^I(t')$ to be less than unity, we can truncate the series up to first order in the interaction Hamiltonian and we can express the propagator above in a much simpler form as (when the system goes from an initial state $|\psi_i\rangle$ to a final state $|\psi_f\rangle$) as
\begin{equation}\label{III.8}
\langle\psi_f|\hat{\mathcal{U}}^I(t,t_i)|\psi_i\rangle\simeq -\frac{i}{\hbar} \int_{t_i}^t dt' \langle \psi_f|\hat{\mathcal{H}}_{int}^I(t')|\psi_i\rangle~.
\end{equation}
We can consider the initial state of the system to be a tensor product state at $t_i=0$ which enables us to write down $|\psi_i\rangle$ as $|\psi_i\rangle=|n_{G_i}\rangle\otimes|n_{P_i}\rangle\otimes |n_{R_i}\rangle=|n_{G_i},n_{P_i},n_{R_i}\rangle$
where $n_{G_i}$ denotes the number of gravitons in the initial state of the system, $n_{P_i}$ denotes the number of photons in the initial state of the system, and $n_{R_i}$ denotes the energy level of the gravitational wave bar detector. Similarly, the final state of the system can be expressed as $|\psi_f\rangle=|n_{G_f}\rangle\otimes|n_{P_f}\rangle\otimes |n_{R_f}\rangle=|n_{G_f},n_{P_f},n_{R_f}\rangle$.
We consider $n_{R_i}=0$ which denotes that the system initially is in a ground state. Then the initial state is simply $|n_{G_i},n_{P_i},0\rangle$. We can now calculate the propagator in eq.(\ref{III.8}) by using the analytical form of the interaction Hamiltonian in the interaction picture. We now consider case for which $n_{R_f}=1$ implying that the detector only jumps a single energy level. In eq.(\ref{III.8}), one can indeed set $t_i=-\infty$ and $t=\infty$ for more clarity while obtaining the result for the transition amplitude and eventually the transition probability. If we now also require that $n_{G_f}\neq n_{G_i}$, the sole non-vanishing contributions to the propagator will come from the interaction Hamiltonian considering the graviton-photon-detector coupling given in eq.(\ref{III.6}). The transition amplitude for the system to go from an initial state $|\psi_i\rangle=|n_{G_i},n_{P_i},0\rangle$ to some final state $|\psi_f\rangle=|n_{G_f},n_{P_f},1\rangle$ then reads
\begin{widetext}
\begin{equation}\label{III.9}
\begin{split}
&\langle\psi_f|\hat{\mathcal{U}}^I(\infty,-\infty)|\psi_i\rangle=-\frac{\pi g_hq_P}{mm_0}\sqrt{\frac{m\hbar\omega}{2m_0\omega_0m_P\Omega_P}}\left[\sqrt{n_{G_i}+1}\sqrt{n_{P_i}}\delta_{n_{G_f},n_{G_i}+1}\delta_{n_{P_f},n_{P_i}-1}\delta\left(\Omega_P-(\omega+\omega_0)\right)\right.\\&\left.-\sqrt{n_{G_i}}\sqrt{n_{P_i}}\delta_{n_{G_f},n_{G_i}-1}\delta_{n_{P_f},n_{P_i}-1}\delta(\omega_0-(\omega+\Omega_P))-\sqrt{n_{G_i}}\sqrt{n_{P_i}+1}\delta_{n_{G_f},n_{G_i}-1}\delta_{n_{P_f},n_{P_i}+1}\delta(\omega-(\omega_0+\Omega_P))\right]
\end{split}
\end{equation}
\end{widetext}
where the unphysical process containing a Dirac delta function $\delta(\omega+\Omega_P+\omega_0)$ has been dropped. 
Depending on the frequency of the graviton, photon, and oscillator, three distinct possibilities are there while the detector jumps from its ground state to higher excited state. Here, we will discuss about two of the more physically involved processes that shall really help in detection of graviton signatures in future generation of graviton detectors.

\noindent \textit{Stimulated emission of gravitons due to absorption of photons with higher frequency:} If the modulated frequency of the photon $\Omega_P$ is equal to that of the combined frequency of the detector and the graviton, then in such a scenario the first term inside of the parenthesis in the analytical expression of the transition amplitude (in eq.(\ref{III.9})) gives a non-vanishing contribution. This scenario depicts that if the detector is in the ground state and the resonance condition is satisfied one single photon will be absorbed while detector jumps to the higher excited state with a simultaneous emission of a graviton particle with frequency $\omega$ such that $\Omega_P>\omega,\omega_0$ and $\Omega_P=\omega+\omega_0$. The transition probability in such a scenario will read $\mathcal{P}_{if}=|\langle\psi_f|\hat{\mathcal{U}}^I(\infty,-\infty)|\psi_i\rangle|^2$ and is given as
\begin{equation}\label{III.10}
\begin{split}
\mathcal{P}_{if}^{\text{Stim.}}=
&\frac{\pi^2\hbar\omega g_h^2q_P^2}{2mm_0^3\omega_0m_P\Omega_P}(n_{G_i}+1)n_{P_i}\delta^2[\Omega_P-(\omega+\omega_0)]
\end{split}
\end{equation}
while $n_{G_f}=n_{G_i}+1$ and $n_{P_f}=n_{P_i}-1$. If $n_{G_i}=0$ and $n_{P_i}=1$, then it simply tells that a photon with high frequency gets converted to a low frequency graviton while the detector jumps a single energy level. Now, the important thing to observe is that the transition probability is proportional to the number of photons in the initial state of the system. If the number of photons can be increased significantly then it will lead to a significant jump in the transition probability acting as pump photons. For the standard detector-graviton coupling case in \cite{OTMGraviton}, it is not possible to increase the number of gravitons arbitrarily as it is beyond the capabilities of the experimental set-up. However, here one can significantly increase the emission probability of the gravitons just by pumping photons in the initial state which indeed is experimentally achievable. The opposite scenario will be that the detector absorbs a low frequency graviton and it jumps down one energy level from its higher excited state while emitting a high frequency photon.

\noindent \textit{Stimulated emission of photons due to absorption of gravitons with higher frequency:} This is process is more interesting. Consider that the frequency of the graviton is higher than the detector as well as the photon field being considered. If the resonance condition $\omega=\omega_0+\Omega_P$ is satisfied in that case the detector jumps to the higher excited state by absorbing a graviton while simultaneously emitting a photon. This indeed is a very interesting physical scenario as the graviton gets converted to a low frequency photon resulting in a visible signature of quantum gravity. One can again calculate the transition probability for this scenario, which comes out to be 
\begin{equation}\label{III.11}
\begin{split}
\mathcal{P}_{if}^{\text{Stim.}}
=&\frac{\pi^2\hbar\omega g_h^2q_P^2}{2mm_0^3\omega_0m_P\Omega_P}n_{G_i}(n_{P_i}+1)\delta^2[\omega-(\omega_0+\Omega_P)]~.
\end{split}
\end{equation}
As before it is evident that increasing the number of photons in the initial state of the system one can substantially increase the transition probability for the model detector system. 

\noindent \textit{Spontaneous graviton emission from photon conversion:} A more interesting scenario is observed while the detector is in its first excited state then if the detector de-excites while absorbing a photon then it will emit a graviton which has the frequency equal to the detector and the photon. The transition probability for the same case reads
\begin{equation}\label{III.12}
\mathcal{P}_{if}=\frac{\pi^2\hbar\omega g_h^2q_P^2}{2mm_0^3\omega_0m_P\Omega_P}(n_{G_i}+1)n_{P_i}\delta^2[\omega-(\omega_0+\Omega_P)]~.
\end{equation}
If now there are no gravitons in the initial state of the system $n_{G_i}=0$ then also the detector will absorb a photon and de-excite while spontaneously emitting a graviton and the transition probability simply reads
$\mathcal{P}_{if}^{\text{Spont.}}=\frac{\pi^2\hbar\omega g_h^2q_P^2}{2mm_0^3\omega_0m_P\Omega_P}n_{P_i}\delta^2[\omega-(\omega_0+\Omega_P)]$
which is again proportional to $n_{P_i}$. It is now thus possible to increase the number of pump photons in the initial state of the system which will significantly increase the spontaneous emission probability of the gravitons. This will be a significant advancement towards already proposed  experimental scenarios for the detection of graviton signatures.

\noindent \textit{Comparison with the semiclassical analogue:} We have very recently proposed a model for detecting gravitational waves using charged Weber bars, place inside of a cavity filled with photons in \cite{OTM}. For the semiclassical model we shall start with the case of stimulated photon emission as a result of absorption of gravitons by a charged Weber bar. The transition probability is given in eq.(\ref{III.11}). From our earlier analysis \cite{OTMGraviton}, we have established that the standard stimulated graviton emission can be mapped directly to its semi-classical analogue where we primarily have investigated the transition due to the interaction Hamiltonian $\frac{g_h}{2mm_0}\hat{p}_h\otimes (\hat{\xi}\hat{\pi}_\xi+\hat{\pi}_\xi\hat{\xi})$. The first step is to consider that the energy carried by the gravitational wave with frequency $\omega$ can be considered as the total energy carried by $n_{G_i}$ number of gravitons each carrying an energy $\hbar \omega$. We then need to use the energy flux relation of a gravitational wave given by $\frac{dE}{dA}=\frac{1}{32\pi G}\int_{-\infty}^{\infty}dt \langle \dot{h}_{ij}^{TT}\dot{h}_{ij}^{TT}\rangle$ with $\langle\cdots\rangle$ denoting a temporal average. Following the methodology in \cite{OTMGraviton} (Eq.(35)), we obtain the relation between the amplitude of the gravitational wave and the number of gravitons $n_{G_i}$ to be
\begin{equation}\label{III.14}
n_{G_i}\hbar\omega=\frac{\pi\omega f_0^2 L^2}{4G}
\end{equation}
where $L=2\pi^2/\omega$ when $c$ is taken to unity. This helps us to reduce the transition probability in eq.(\ref{III.11}) as
\begin{equation}\label{III.15}
\begin{split}
\mathcal{P}_{if}^{\text{QG}}=&\frac{\pi^2\hbar\omega g_h^2q_P^2}{2mm_0^3\omega_0m_P\Omega_P}n_{G_i}(n_{P_i}+1)\delta^2(\omega-\omega_0-\Omega_P)\\
=&\frac{\pi^2\omega^2 q_P^2f_0^2}{4m_0\omega_0m_P\Omega_P}(n_{P_i}+1)\delta^2(\omega-\omega_0-\Omega_P)=\mathcal{P}^{\text{GW}}_{if}
\end{split}
\end{equation}
which is exactly equal to the transition probability obtained in eq.(\ref{III.11}) of \cite{OTM}. This gives an excellent analogy between the semiclassical and the quantum gravity pictures and establishes the correctness of the model being considered here. However, one can never reproduce the spontaneous emission of the gravitons case is unique to the quantum gravity treatment. With this analogy, we can finally proceed to propose and experimental model based on this proposal.

\noindent \textbf{\textit{Experimental proposal for detecting emitted photons:}} We start with the model where a high frequency graviton is absorbed by the detector while it jumps to a higher excited state while simultaneously emitting a low frequency photon and the condition for occurrence of this phenomena is simply given by the resonance condition $\omega=\omega_0+\Omega_P$. In a realistic measurement scenario the total measurement time can be considered to be $\tau$ which allows for expressing the Dirac delta function as
$\delta(\omega-\omega_0-\Omega_P)\rightarrow\frac{1}{2\pi}\int_{-\frac{\tau}{2}}^\frac{\tau}{2}dt~e^{i(\omega-\omega_0-\Omega_P)}=\frac{1}{\pi(\omega-\omega_0-\Omega_P)}\sin\left[(\omega-\omega_0-\Omega_P)\frac{\tau}{2}\right]$ and when the resonance condition is satisfied it is possible to replace the Dirac delta function as $2\pi\delta(\omega-\omega_0-\Omega_P)\rightarrow \tau$. The transition rate can then be obtained simply by the expression (with a proper dimensional reconstruction) as $\Gamma_{if}=\frac{1}{\tau}\mathcal{P}_{if}=\frac{\pi^2 q^2 G}{\epsilon_0 c^2 m_0\omega_0\Omega_P L^6}n_{G_i}\hbar\omega(n_{P_i}+1)\delta(\omega-\omega_0-\Omega_P)$ where $\epsilon_0$ denotes the permittivity of free space and $L$ is the side of the box where the gravitational as well as electromagnetic wave is being quantized. In order to obtain the transition rate, we need to fix the parameter values given by $L=1$ m, $q=1$ C, $m_0=10^{-6}$ kg, $\omega=1000$ Hz, $\omega_0=900$ Hz, $\Omega_P=100$ Hz, and $\tau=10^5$ sec. This allows us to obtain the transition rate exactly at the resonance point to be ($\delta(\omega-\omega_0-\Omega_P)\rightarrow \frac{\tau}{2\pi}$) $\Gamma_{if}\sim 10^{-41} n_{G_i}(n_{P_i}+1)$ $\text{sec}^{-1}$ which is extremely small for very low initial number of gravitons as well as photons. For a gravitational wave with amplitude $f_0=10^{-21}$ one can find out that the number of gravitons will roughly be of the order of $n_{G_i}^{\text{max}}\sim 10^{27}$ which gives the transition rate to be of the order of $\Gamma_{if}\sim 10^{-14} (n_{P_i}+1)$ $\text{sec}^{-1}$. The important thing to observe is that the spontaneous emission probability becomes then very small and almost impossible to detect. The first way to enhance the effective transition rate is to consider oscillators with very small length such that the effective quantization volume can be made small. One way is to use LC oscillators with charged plated so that it mimics our analytical model. The other way is to pump the optical cavity using pumping mechanisms such that a large number of photons accumulate in the initial state of system leading to an overall effective rise in the transition rate.

\begin{widetext}

\begin{figure}
\begin{center}
\includegraphics[scale=0.875]{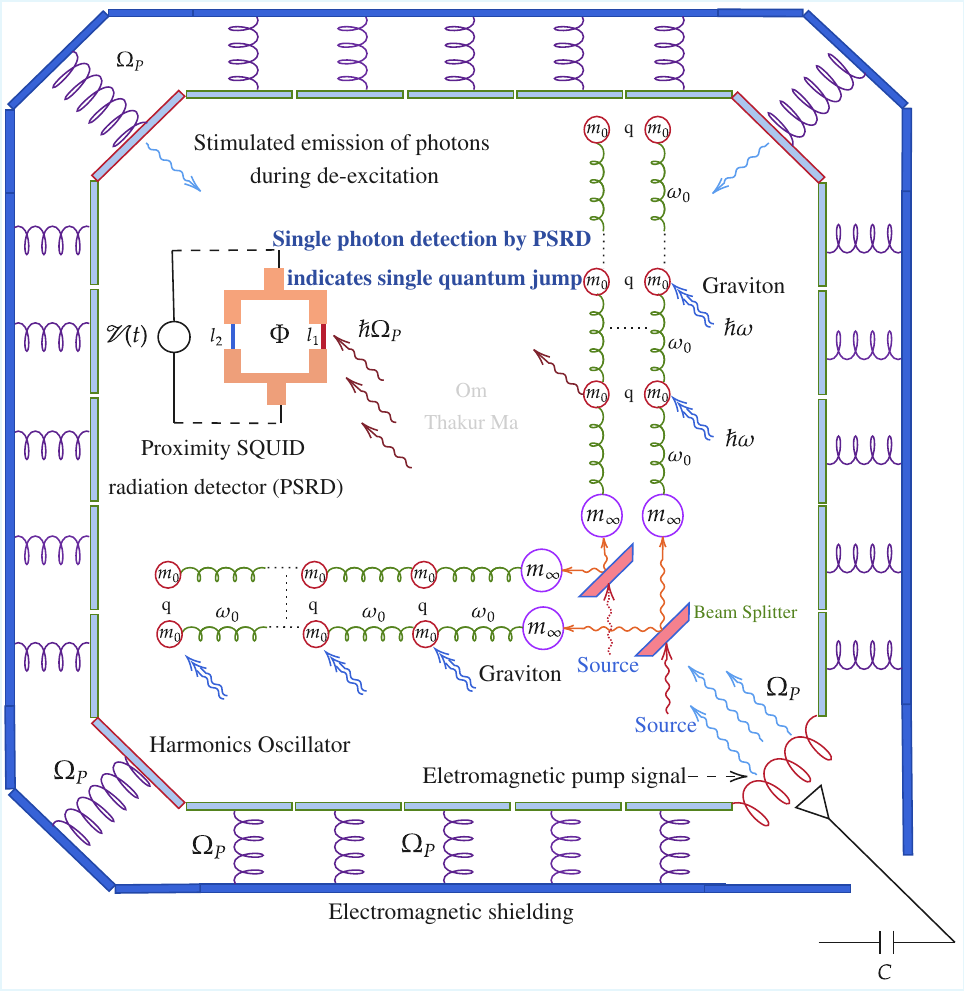}
\caption{A schematic graviton detector model (not to scale) based on the conversion of gravitons into lower frequency photons via an array of fully coherent harmonic oscillators. The measurement of single photon emissions via a proximity superconducting quantum interference device radiation detector (PSRD) indicates the indirect detection of absorption of a single graviton.\label{Detector_OTM}}
\end{center}
\end{figure}

\end{widetext}

\noindent A simple effective model (a schematic diagram) for graviton detection can be seen in Fig.(\ref{Detector_OTM}). Here, the walls of a optical cavity is made of small vibrating oscillators while from one an electromagnetic signal with oscillation frequency $\Omega_P$ is pumped inside of the cavity. This allows for the detectors at the walls to get excited and during de-excitation they emit photons with frequency $\Omega_P$ ($\Omega_P\sim \omega_P$) inside of the electromagnetically shielded cavity creating a large number of photons in the initial state of the system. This allows for a huge gain in $n_{P_i}$. The next step is to set-up an array of identical oscillators. If $N$ number of oscillators are connected in an array an effective amplitude gain of $\mathcal{A}_{if}\rightarrow N\mathcal{A}_{if}$ is obtained resulting an overall $N^2$ gain in the transition probability. If similar arrays are placed parallel and perpendicular to the initial array with a full coherence generation using beam-splitters, a further enhancement can be obtained depending on the number of such $L$ shaped set-ups as can be seen from Fig.(\ref{Detector_OTM}). If $n$ number of $L$ shaped set-ups are added an overall gain of $\mathcal{P}_{if}\rightarrow nN^2\mathcal{P}_{if}$ can be observed. Each of the oscillators are charged and carrying a charge of $q$ C where the oscillators have a fundamental frequency equal to $\omega_0$ Hz. All the detectors are cooled so that the entire system is in its ground state. The cooling is not shown in the figure. If the graviton carrying an energy $\hbar \omega$ interacts with the oscillators then each of the oscillator jumps to its first excited state while emitting a single photon with effective frequency $\Omega_P$ such that the resonance condition $\omega=\omega_0+\Omega_P$ gets satisfied. The emitted photons are then detected by a ``Proximity Superconducting Quantum Interference Device (SQUID) Radiation Detector" or a PSRD \cite{PSRD}. In this scenario, we suggest the use of single photon detection scenario presented in \cite{PSRD}. The PSRD (as can be seen from the figure as well)  is constructed using two Superconductor-Normal metal-Superconductor (SNS) Josephson junction weak-links where the orange bars are denoting the superconductors and $l_1$ and $l_2$ denotes the length of the weak-links. An external magnetic flux $\Phi$ pierces this SQUID configuration and a single photon absorption generated a time dependent pulse in the measuring interferometer (a Voltage change $\mathcal{V}(t)$) resulting in a single photon detection.  As $\hbar\Omega_P$ denotes the total energy of a single emitted photon, an overall Voltage change divided by the voltage change due to single photon gives the total number of photon emissions which equals the total number of quantum jumps made by the oscillators. It is important there will always be a constant voltage measurement for the pumped photons and as a result any change above that value will actually count towards quantum jumps by the \textbf{transducers} or the array of harmonic oscillators. Each quantum jump actually signifies absorption of a single graviton by the harmonic oscillator and hence, this experiment will actually give us an indirect way of seeing the quantum ripples of spacetime. If $n_{G_i}=10^{27}$, even then just by a pumping of $n_{P_i}\sim 10^{14}$ number of photon, we can observe one photon emission per second which is indeed a very promising result without using an array of such oscillators.

\noindent In a realistic scenario, however, $n_{G_i}$ can be very small such that $n_{G_i}\sim 10^{12}$ for a such a set up a the transition rate becomes $\Gamma_{if}\sim 10^{-29}(n_{P_i}+1)$ $\text{sec}^{-1}$. A pumping up to $n_{P_i}\sim 10^{22}-10^{24}$ can be achieved which allows for a transition rate of  $\Gamma_{if}\sim 10^{-7}-10^{-5}$ $\text{sec}^{-1}$. Using a $10-100$ array oscillator set up with $10$ $L$ shaped fully coherent systems, one can achieve an overall gain of $10^3-10^5$ gain in the transition rate giving it a more measurable value of $\Gamma^{\text{Eff}}_{if}\sim 10^{-4}-1$ $\text{sec}^{-1}$ where probabilities from the individual detectors remain the same.

\noindent \textbf{\textit{Experimental proposal for detecting spontaneous emission of gravitons:}} Although the previous set-up allows a promising way of detecting gravitons, it heavily relies on the fact that a weak classical perturbation cannot result in consistent quantum jumps. It is although partially true, one can propose a more robust graviton detector just by using the spontaneous graviton emission scenario. After excitation when the detector de-excites it absorbs a single photon emitting a single graviton spontaneously. Hence, the overall drop of the voltage will indicate the loss of photons in the cavity. To achieve this set-up the entire set-up needs to be excited to its first excited state. This can be achieved by temperature control and fixing the overall cavity temperature to a value such that $T\sim\frac{3\hbar \omega_0}{2k_B}$ which allows for the population of the state via thermal excitation. Then during de-excitation the oscillators absorb a photon with energy $\hbar\Omega_P$ while emitting a graviton with energy $\hbar\omega$. The spontaneous emission scenario is very difficult to observe as even with a 100 oscillator- 10 $L$ shaped set-up the transition rate gives an estimated output to be $\Gamma^{\text{Spont.}}_{if}\sim 10^{-36} n_{P_i}$ $\text{sec}^{-1}$. The only way to increase the transition probability for the spontaneous emission is through electromagnetic pumping and achieving a total $n_{P_i}$ value of the order of $10^{30}-10^{33}$ which is very difficult to attain experimentally. Now,  the scenario may change significantly if the quantization volume can be significantly decreased and made of the order of $V\sim 10^{-6}$ $\text{m}^3$ which allows for a overall gain of $10^{12}$ in the transition rate as $\mathcal{P}_{if}^{\text{Spont.}}\propto\frac{1}{V^2}$. Then the transition rate bumps to $\Gamma^{\text{Spont.}}_{if}\sim 10^{-24} n_{P_i}$ $\text{sec}^{-1}$ which allows for a respectable pumping of the order of $n_{P_i}\sim 10^{22}-10^{25}$ matching the transition rate for the previous case. Such LC oscillators incorporate the graviton effects as the time dependent perturbation changes the effective separation in the capacitor plates and changes the geometry of the inductors resulting in a more experimentally complex but feasible graviton detector. In case of stimulated emission scenario the cavity walls can be created using two layers of oscillators where the second or the inner most layer will have oscillators with frequency $\omega$. This walls will absorb the emitted the gravitons to execute the quantum jumps. The simultaneous quantum jump from the cavity walls and voltage drop in the PSRD will confirm detection of gravitons by our proposed detector model and also confirm the existence of the quantum gravitational Gertsenshtein effect where the array of harmonic oscillators work as an effective transducer.

\noindent \textit{\textbf{Circumventing Dyson's argument on the nondetectability of gravitons:}} Freeman Dyson \cite{Dyson} has long argued that it is impossible to detect a graviton in a realistic experimental set-up. Here, we shall primarily consider the pure Gertsenshtein conversion probability that Dyson calculated in \cite{Dyson}. Dyson showed that for photon travelling a distance $D$ through uniform magnetic field $B$, the probability of the photon to emerge as a graviton has a probability $\mathcal{P}_{\text{Dyson}}=\frac{GB^2D^2}{4c^4}$ and for the Gertsenshtein process to be successful $DB^2\omega\leq 10^{43}$ which makes the transition probability to be $\mathcal{P}_{\text{Dyson}}\leq 10^{36}B^{-2}\omega^{-2}$. For, $B=10^5$ Gauss and $D=10^{13}$ cm the maximum transition probability becomes $10^{-14}$. This probability decreases further for large $B$ (as the propagation distance $D$ needs to get smaller for larger $B$ for the process to succeed). Now our model has some fundamental difference from the standard Gertsenshtein models as our analysis shows that the detector acts as a transducer for the gravitons getting converted to photons or vice-versa. The next important thing is that our model is proposing to detect the change in the voltage in the PSRD due to the converted photons and as a result our proposal is not based on directly detecting gravitons. On the other hand the electromagnetic pumping along with arrays of tiny quantum detectors allows for a significant amplification of the conversion probability and for the spontaneous emission scenario $\mathcal{P}_{if}^{\text{Eff.}}\propto nN^2n_{P_i} G$ where all $n$, $N$, and $n_{P_i}$ are experimentally controllable. The transition probability in such a case can be in the range $\mathcal{P}_{if}=\tau \Gamma_{if}\sim 10^{-2}-1$ for the best case scenario and is many orders of magnitude higher compared to the case investigated in \cite{Dyson}. Even for a electromagnetic pumping $n_{P_i}\sim 10^{16}$ (which way lower than what is achievable in an actual experimental set-up) the transition probability is $\mathcal{P}_{if}\sim 10^{-8}\gg \mathcal{P}_{\text{Dyson}}$. This simple calculation shows that our model allows for a realistic graviton detection scenario which allows for bypassing Dyson's original argument for the nondetectability of gravitons which is a direct consequence of the trilinear interaction with a fundamentally different physical scenario.

\noindent \textbf{\textit{Discussion and Conclusion:}} In this work, we propose a novel graviton detection scenario using the quantum gravitational analogue of the Gertsenshtein effect where a high frequency graviton is absorbed by a harmonic oscillator or a Weber bar while it jumps to a higher excited state while simultaneously emitting a single photon with lower frequency and vice-versa. We start via considering a charged Weber bar in presence of a dynamical electromagnetic field. The charge allows for the detector to interact with the electromagnetic field. The total action of the system is then obtained by combining the Einstein-Hilbert, minimally coupled Resonant detector action, and Maxwell's equation in curved background. We have then obtained the total Hamiltonian after which we have raised the phase space variables to operator status and implemented suitable canonical commutation relation to quantize the model system. We observe that the interaction Hamiltonian is composed of known graviton-detector, graviton-photon (QG Gertsenshtein), photon-detector, and the novel trilinear graviton-photon-detector coupling terms. We then investigate the trilinear term which allows for the quantum mechanical detector to act as transducer which depending on the resonance condition allows conversion of gravitons into photons and vice-versa while it excites or de-excites from its initial energy level. We primarily look for the case when a high frequency graviton is absorbed by the detector while it excites to a higher energy level with the simultaneous emission of a lower frequency photon. While the detector de-excites it emits a high frequency graviton by absorbing a low frequency photon. For the photon emission case, we have find out that it can be mapped exactly to its semiclassical analogue discussed earlier in \cite{OTM}. We have then proposed a graviton detector based on the theoretical analysis of the model. For the model, we propose the use of a cavity-QED set up where an array of full coherent charged harmonic oscillators are placed. The cavity is made of oscillating walls for allowing electromagnetic pumping through stimulated emission of photons. Then a PSRD is installed to detect single photon emission due to quantum jumps by the detector as a result of absorbing high frequency gravitons. The spontaneous emission scenario, however, can also be measured using small LC oscillators and two-layer cavity walls with oscillators installed for allowing electromagnetic pumping and graviton induced quantum jumps. As the detector is designed to work in $10^3$ Hz frequency range, a simultaneous detection of a gravitational wave with the LIGO with our proposed model will confirm the detection of gravitons in this set-up where it actually possible to indirectly visualize the gravitons through photon emission. 
Finally, we compare our transition probability to the Dyson's original argument on the non-detectability of gravitons and analytically prove that our model circumvents the Dyson-argument and our transition probability is several orders of magnitude higher than the model proposed in \cite{Dyson}.
\smallskip

\noindent\textbf{Acknowledgement:} S. Sen thanks Dr. Igor Pikovski for an email correspondence over the manuscript.

\end{document}